\documentclass[aps,prd,twocolumn,superscriptaddress,nofootinbib]{revtex4-2}

\usepackage{amsmath}
\usepackage{amssymb}
\usepackage{graphicx}
\usepackage{hyperref}

\begin{document}
	
	\title{Radiative decay of $\chi_{c1}$ states in effective Lagrangian approach}
	
	\author{Thanat~Sangkhakrit}
	\email{thanat.san@kmitl.ac.th}
	\affiliation{School of Physics and Center of Excellence in High Energy Physics and Astrophysics, Suranaree University of Technology, Nakhon Ratchasima 30000, Thailand}
	\affiliation{KOSEN Institute of King Mongkut's Institute of Technology Ladkrabang, Bangkok 10520, Thailand}
	
	\author{Attaphon~Kaewsnod}
	\affiliation{School of Physics and Center of Excellence in High Energy Physics and Astrophysics, Suranaree University of Technology, Nakhon Ratchasima 30000, Thailand}
	
	\author{Ayut~Limphirat}
	\affiliation{School of Physics and Center of Excellence in High Energy Physics and Astrophysics, Suranaree University of Technology, Nakhon Ratchasima 30000, Thailand}
	
	\author{Khanchai~Khosonthongkee}
	\affiliation{School of Physics and Center of Excellence in High Energy Physics and Astrophysics, Suranaree University of Technology, Nakhon Ratchasima 30000, Thailand}
	
	\author{Warintorn~Sreethawong}
	\affiliation{School of Physics and Center of Excellence in High Energy Physics and Astrophysics, Suranaree University of Technology, Nakhon Ratchasima 30000, Thailand}
	
	\author{Nopmanee~Supanam}
	\affiliation{Department of Physics, Faculty of Science, Srinakharinwirot University, Bangkok 10110, Thailand}
	
	\author{Yupeng~Yan}
	\email{yupeng@g.sut.ac.th}
	\affiliation{School of Physics and Center of Excellence in High Energy Physics and Astrophysics, Suranaree University of Technology, Nakhon Ratchasima 30000, Thailand}
	
	%\date{\today}
	
	\begin{abstract}
	The $\chi_{c1}(3872)$ state, first observed by the Belle collaboration with its quantum numbers identified as $J^{PC} = 1^{++}$, has been the subject of extensive research due to its intriguing properties. 
	Several theoretical interpretations have been proposed to explain its unique characteristics, including the $\chi_{c1}(2P)$ assignment, a molecular $\bar{D}^{*}D/\bar{D}D^{*}$ configuration, a coupled-channel framework incorporating $c\bar{c}$ and di-meson degrees of freedom, and the compact tetraquark hypothesis. 
	However, challenges remain in reconciling its mass coincidence with the threshold and the observed isospin violation within both the pure $c\bar{c}$ and compact tetraquark models.
	In this study, we examine the radiative decays of the $\chi_{c1}(1P)$ and $\chi_{c1}(3872)$ states in an effective field theory framework, incorporating triangle loops of $D$ and $D^{*}$ mesons. 
	The model parameters are calibrated based on the observed branching fraction of the radiative decay mode $\chi_{c1}(1P) \to J/\psi \gamma$. 
	Utilizing these fixed parameters, we predict the branching fractions $R_{\chi_{c1}(3872) \to J/\psi \gamma} \sim 10^{-1}$ and $R_{\chi_{c1}(3872) \to \psi(2S) \gamma} \sim 10^{-2}$, and the relative fraction $\mathcal{R}_{\Psi\gamma} \approx 0.109$.
	The work supports the argument that the $\chi_{c1}(3872)$ is unlikely a $c\bar{c}$ state.

	\end{abstract}
	
	\maketitle
	
	\section{Introduction}

	The exotic state $\chi_{c1}(3872)$ was first discovered by the Belle collaboration in 2003 through the decay process $B^{+} \to J/\psi \pi^{+} \pi^{-} K^{+}$ \cite{Belle:2003nnu}. 
	Since then, its properties have been extensively investigated in various collision modes by the Belle \cite{Belle:2003eeg,Belle:2006olv,Belle:2008fma,Belle:2011vlx,Belle:2011wdj,Belle:2015qeg,Belle:2022puc,Belle:2023zxm},
	BABAR \cite{BaBar:2004iez,BaBar:2004oro,BaBar:2006fjg,BaBar:2007cmo,BaBar:2007ixp,BaBar:2008flx,BaBar:2008qzi,BaBar:2010wfc},
	CDF \cite{CDF:2003cab,CDF:2005cfq,CDF:2006ocq,CDF:2009nxk},
	D0 \cite{D0:2004zmu},
	ATLAS \cite{ATLAS:2016kwu},
	CMS \cite{CMS:2013fpt,CMS:2020eiw},
	and LHCb collaborations \cite{LHCb:2011zzp,LHCb:2013kgk,LHCb:2014jvf,LHCb:2015jfc,LHCb:2016zqv,LHCb:2019imv,LHCb:2020fvo,LHCb:2020sey,LHCb:2020xds,LHCb:2021ten,LHCb:2022jez,LHCb:2022oqs,LHCb:2024bpb}. 
	After significant efforts, its quantum numbers were confirmed as $J^{PC} = 1^{++}$.
	Due to its intriguing properties, including the closeness of its mass to the $D^{0}\bar{D}^{*0}$ threshold, isospin violation in the decay modes $\chi_{c1}(3872) \to J/\psi \pi^{+}\pi^{0}\pi^{-}$ and $\chi_{c1}(3872) \to J/\psi \pi^{+} \pi^{-}$, and a large coupling to the $D^{0}\bar{D}^{*0}$ system, this state has garnered significant attention.

	To date, various theoretical interpretations have been proposed to investigate the radiative decay properties of $\chi_{c1}(3872)$, particularly the radiative branching fraction $R_{\chi_{c1}(3872) \to J/\psi \gamma}$. 
	In \cite{Barnes:2003vb,Kalashnikova:2010hv,Yu:2023nxk}, $\chi_{c1}(3872)$ is hypothesized to be the first radial excitation state, $\chi_{c1}(2P)$. 
	The branching fraction $R_{\chi_{c1}(3872) \to J/\psi \gamma}$ predicted by this model is about two orders of magnitude larger than the observed fraction \cite{Yu:2023nxk}.
	The molecular $\bar{D}^{*}D / \bar{D}D^{*}$ configuration is considered in \cite{Tornqvist:2004qy,Liu:2008fh,Liu:2009qhy,Li:2012cs,Guo:2013zbw,Guo:2014taa,Chen:2024xlw} due to the mass coincidence and its large coupling to the $D^{0}\bar{D}^{*0}$ system. 
	The branching fraction $R_{\chi_{c1}(3872) \to J/\psi \gamma}$ predicted by this model is of the same order of magnitude as the observed fraction \cite{Chen:2024xlw}. 
	The coupled-channel framework, which incorporates $c\bar{c}$ and di-meson degrees of freedom, is employed in \cite{Braaten:2003he,Kalashnikova:2005ui,Barnes:2007xu,Li:2009ad,Ortega:2009hj,Baru:2010ww,Cardoso:2014xda,Yamaguchi:2019vea,Takeuchi:2021cnp}. 
	This model predicts that the branching fraction $R_{\chi_{c1}(3872) \to J/\psi \gamma}$ is about one order of magnitude higher than the observed fraction \cite{Cardoso:2014xda}.
	The compact tetraquark model is explored in \cite{Maiani:2004vq,Maiani:2007vr}.
	Nevertheless, several challenges persist, particularly its mass coincidence with the threshold and the observed isospin violation, which are difficult to reconcile within the pure $c\bar{c}$ and compact tetraquark paradigms.
	The recent experimental observation at LHCb \cite{LHCb:2024tpv} challenges the interpretation of $\chi_{c1}(3872)$ as a purely molecular $D^{0}\bar{D}^{*0} + \bar{D}^{0}D^{*0}$ state.
	This finding strongly suggests the presence of a sizeable compact charmonium or tetraquark component within the $\chi_{c1}(3872)$ state.

	The upgrade of CEBAF's electron beam energy to 22 GeV presents an excellent opportunity to address these challenges. 
	This advancement enables precise measurements of the energy dependence of production rates, facilitating a detailed investigation into the production mechanisms of the $\chi_{c1}(1P)$ and $\chi_{c1}(3872)$ states in $\gamma^* p$ interactions \cite{Accardi:2023chb}. 
	This will provide not only a better understanding of the nonperturbative dynamics of QCD but also insights into the internal structures of these states.

	In this study, we examine the radiative decays of the $\chi_{c1}(1P)$ and $\chi_{c1}(3872)$ states using an effective Lagrangian approach, incorporating triangle loops of $D$ and $D^*$ mesons. 
	First, the model parameters are calibrated using the observed branching fraction of the radiative decay $\chi_{c1}(1P) \to J/\psi \gamma$. 
	With these parameters fixed, we predict the branching fractions for the decay channels $\chi_{c1}(3872) \to J/\psi \gamma$ and $\chi_{c1}(3872) \to \psi(2S)\gamma$, as well as their relative fraction.

	The content of this paper is organized as follows: In Sec.~\ref{EFL}, the effective Lagrangians and form factors for our calculation are introduced. Next, we calculate the radiative branching fraction for $\chi_{c1} \to \Psi \gamma$ in Sec.~\ref{fraction}. The calibration of the model parameters using the observed fraction of the decay mode $\chi_{c1}(1P) \to J/\psi \gamma$ and predictions for the fractions of the decay channels $\chi_{c1}(3872) \to J/\psi \gamma, \psi(2S) \gamma$ are presented in Sec.~\ref{result}. Finally, a summary of this work is given in Sec.~\ref{sum}.

	\section{Effective Lagrangians}
	\label{EFL}

\begin{figure}[h!]
		\begin{center}
		\includegraphics[scale=0.09]{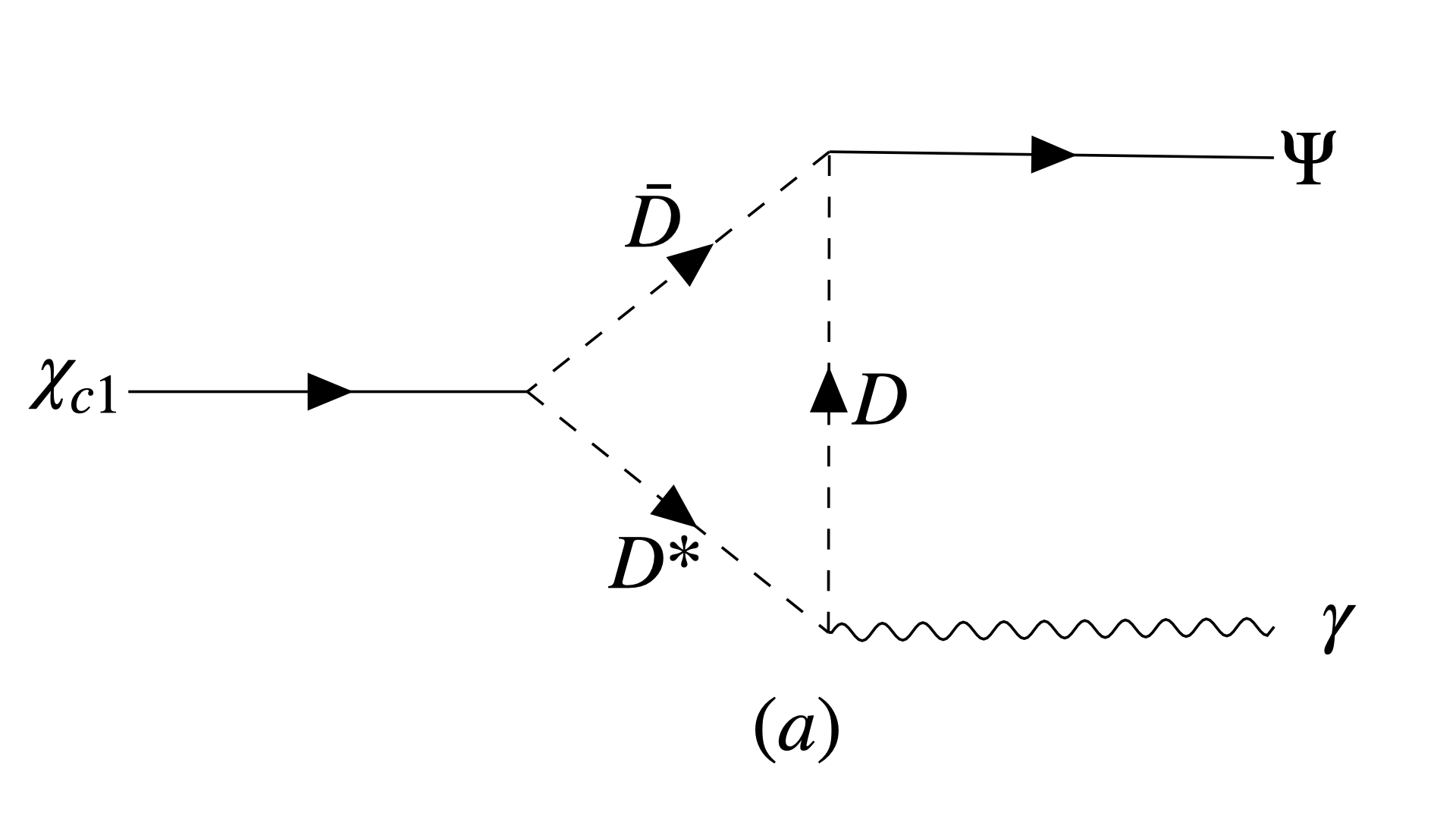} \\
		\includegraphics[scale=0.09]{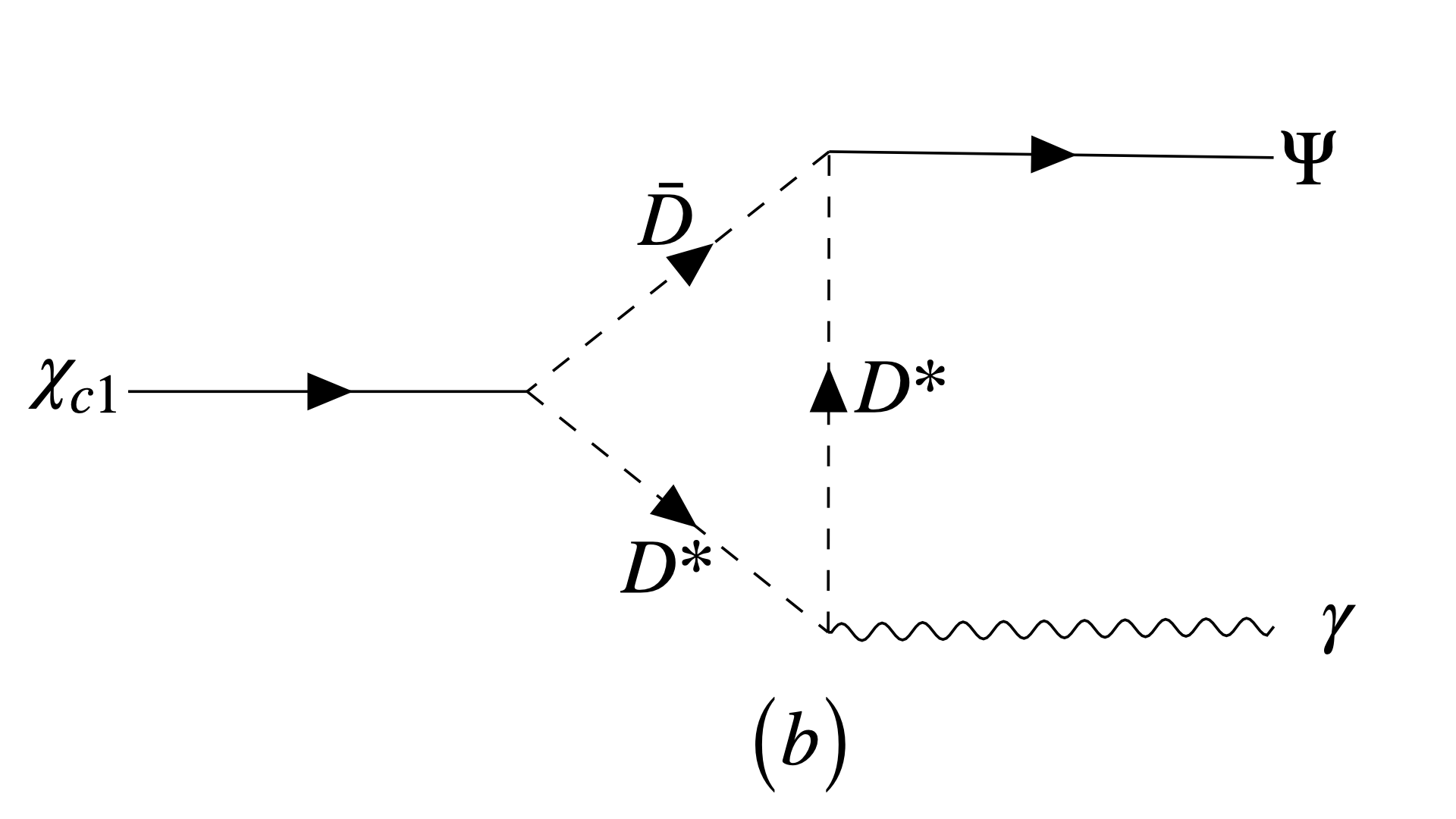} \\
		\includegraphics[scale=0.09]{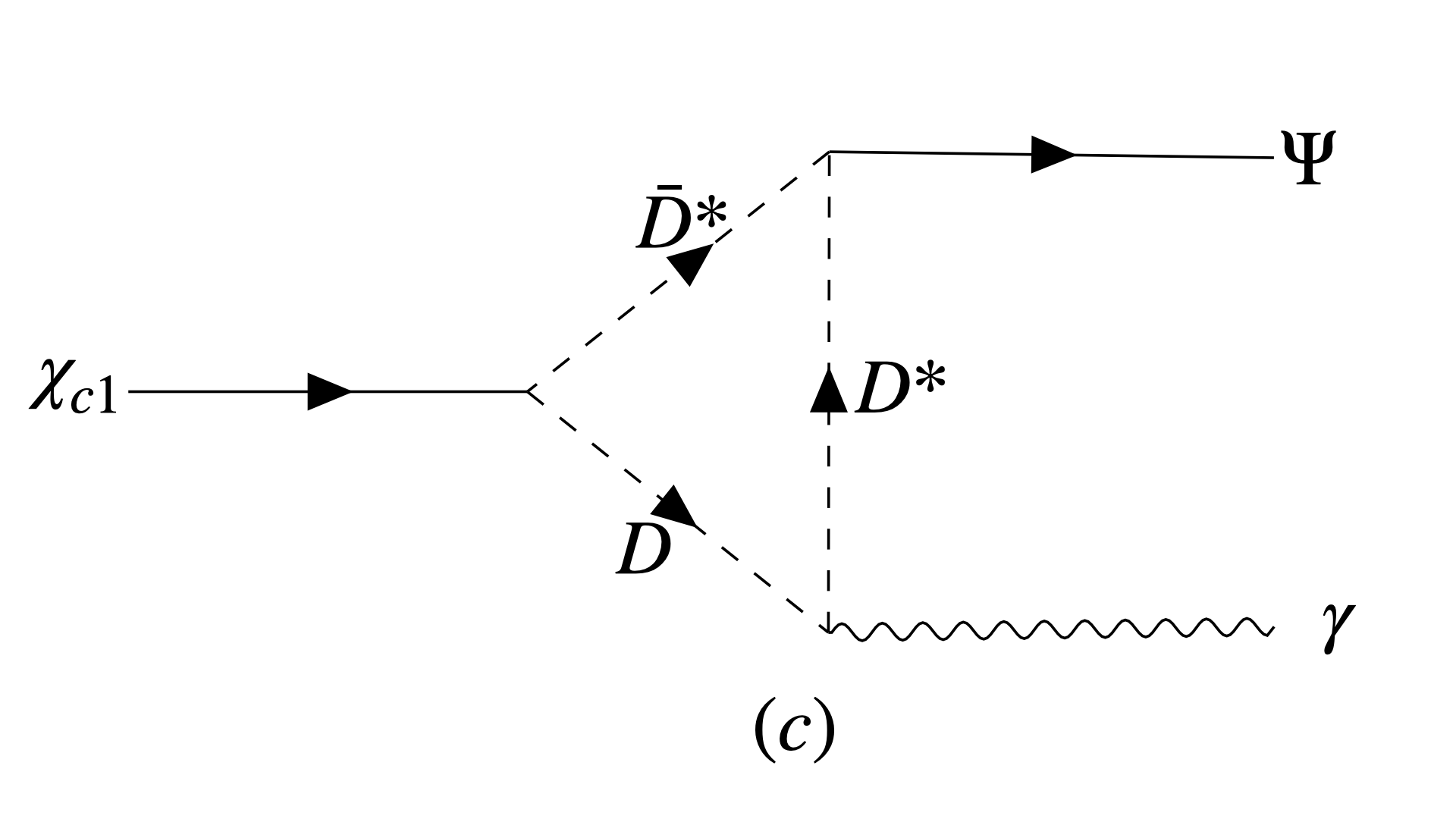}
		\end{center}
		\caption{Feynman diagrams of the radiative decay mode $\chi_{c1} (p_1,\lambda_{1}) \to J/\psi (p_{2},\lambda_{2}) \gamma (p_{3},\lambda_{3}) $ with triangle loops of $D$ and $D^*$ mesons.}
		\label{diagram}
	\end{figure}

	The lowest order diagrams for the radiative decay process $\chi_{c1} \to \Psi \gamma$ with $\chi_{c1} = \left\lbrace \chi_{c1}(1P), \chi_{c1}(3872) \right \rbrace$ and $\Psi = \left\lbrace J/ \psi, \psi(2S) \right\rbrace$ are displayed in Fig.~\ref{diagram}.
	
	Effective Lagrangians in our study are written as \cite{Meng:2007cx,Dong:2008gb}
	\begin{align}
	\mathcal{L}_{\chi DD^{*}} &= g_{\chi DD^{*}} \chi^{\mu} \left( D_{\mu}D^{\dagger} - D D^{\dagger}_{\mu} \right), \nonumber \\
	\mathcal{L}_{\Psi DD} &= i g_{\Psi DD} \Psi_{\mu}  \left( \partial^{\mu} D D^{\dagger} - D \partial^{\mu} D^{\dagger} \right),  \nonumber \\
	\mathcal{L}_{\gamma DD^{*}} &= \frac{e}{4} g_{\gamma DD^{*}} D \epsilon^{\mu \nu \alpha \beta} F_{\mu \nu} D^{\dagger}_{\alpha \beta}, \nonumber \\	
	\mathcal{L}_{\Psi D D^{*}} &= -g_{\Psi D D^{*}} \epsilon^{\mu \nu \alpha \beta} \partial_{\mu} \Psi_{\nu} \left( D \partial_{\alpha} D^{\dagger}_{\beta} + \partial_{\alpha}D_{\beta} D^{\dagger} \right), \nonumber \\
	\mathcal{L}_{\gamma D^{*}D^{*}} &= -i e A^{\mu} \left( D^{\nu} D^{\dagger}_{\mu \nu} - D_{\mu \nu}D^{\nu \dagger} \right) , \nonumber \\
	\mathcal{L}_{\Psi D^{*}D^{*}} &=  -i g_{\Psi D^{*}D^{*}} \left\lbrace \Psi^{\mu}\left( D_{\nu}\partial_{\mu}D^{\nu \dagger} - \partial_{\mu}D_{\nu}D^{\nu \dagger} \right) \right. \nonumber \\ 
	&+ \left. \Psi^{\nu}\partial_{\mu}D_{\nu}D^{\mu \dagger} - \Psi_{\nu}D^{\mu}\partial_{\mu}D^{\nu \dagger} \right\rbrace, \label{L}	
	\end{align}
	with $F_{\mu \nu} = \partial_{\mu}A_{\nu}-\partial_{\nu}A_{\mu}$ and $D_{\mu \nu} = \partial_{\mu}D_{\nu}-\partial_{\nu}D_{\mu}$.
	
	For the $\chi_{c1}D D^{*}$ coupling, the coupling constants $g_{\chi D D^{*}}$ are derived from the $c\bar{c}$ picture \cite{Meng:2007cx}. 
	The value of the coupling constant $g_{\chi D D^{*}}$ is then summarized as
	
	\begin{equation}
		g_{\chi DD^*} =
		\begin{cases}
		21.52 \text{ GeV} & \text{with  } \chi = \chi_{c1}(1P), \\
		23 \text{ GeV} & \text{with  } \chi = \chi_{c1}(3872).
		\end{cases}
	\end{equation}
	
	The coupling constants for $\Psi$ mesons are determined using the vector meson dominance (VMD) mechanism and heavy quark symmetry relations \cite{Meng:2007cx,Chen:2024xlw}. 
	The corresponding coupling constants are summarized in Table.~\ref{psi coupling}. 
	The coupling constant $g_{\gamma D D^{*}} = 2 \text{ GeV}^{-1}$ is extracted from the radiative decay mode of $D^{*0}$ \cite{Dong:2008gb}.
	
	\begin{center}
		\begin{table}[h]
			\begin{tabular}[c]{|c|c|c|}
				\hline
				Coupling constant & $\Psi = J/\psi$ & $\Psi = \psi(2S)$ \\
				\hline
				$g_{\Psi D D}$ & $8$ & $12.39$ \\
				$g_{\Psi D D^{*}}$ & $4.3 \text{ GeV}^{-1}$ & $3.49 \text{ GeV}^{-1}$ \\
				$g_{\Psi D^{*} D^{*}}$ & $8$ & $13.33$ \\
				\hline
			\end{tabular}
			\caption{Coupling constants for $\Psi$ mesons}
			\label{psi coupling}
		\end{table}
	\end{center}
	
	To account for the finite size of hadrons, we introduce a form factor in the monopole form:
	\begin{equation}
		F(q^{2},m) = \frac{\Lambda^{2}-m^{2}}{\Lambda^{2}-q^{2}}, 
		\label{FF}
	\end{equation}
	where $\Lambda$, $m$, and $q^{2}$ represent the cutoff parameter, the mass of the exchanged particle, and the square of the momentum transfer, respectively. In this form factor, the cutoff parameter $\Lambda$ is typically set a few MeV above the mass of the exchanged meson to ensure proper suppression of short-distance effects. 
	To achieve this, we parametrize $\Lambda$ as \cite{Meng:2007cx,Liu:2006df}:
	\begin{equation}
		\Lambda = m + \alpha \Lambda_{\text{QCD}},
	\end{equation}
	where $m$ is the mass of the exchanged meson, and $\Lambda_{\text{QCD}} \approx 0.2 \text{ GeV}$.

	\section{Branching fraction of the radiative decay $\chi_{c1} \to \Psi \gamma$}
	\label{fraction}
	
	By using the Lagrangians in Eq. (\ref{L}) and the form factor in Eq. \ref{FF}, the Feynman amplitudes for the triangle loop diagrams in Fig. \ref{diagram} are written as
	\begin{widetext}
		\small
		\begin{align}
			\mathcal{M}_{a} &= \int \frac{d^{4}q_1}{\left( 2 \pi \right)^{4}} \left( g_{\chi DD^{*}}g_{\psi D D} e g_{\gamma D D^{*}} \epsilon^{\alpha \beta \rho \sigma} \epsilon_{1}^{\mu} \epsilon^{*}_{2\nu}\epsilon^{*}_{3\beta} \right) p_{3 \alpha} \left(p_{1}-q_{1}\right)_{\rho} \left(2q_{1}-p_{2}\right)^{\nu} 
			\left(-g_{\mu \sigma}+\frac{\left(p_1 - q_1\right)_{\mu}\left(p_1-q_1\right)_{\sigma}}{m_{D^{*}}^{2}}\right) \nonumber \\ 
			&\times \left(\frac{1}{q_{1}^{2}-m_{D}^{2}}\right)  \left(\frac{1}{\left(p_1 - q_1 \right)^{2}-m_{D^{*}}^{2}}\right) \left(\frac{1}{\left(p_2 - q_1 \right)^{2}-m_{D}^{2}}\right) 
			F(m_{D},q_{1}^{2}) F(m_{D^{*}},\left(p_{1}-q_{1}\right)^{2}) F(m_{D},\left(p_{2}-q_{1}\right)^{2}), \nonumber \\
			\mathcal{M}_{b} &= \int \frac{d^{4}q_1}{\left( 2 \pi \right)^{4}} \left( g_{\chi D D^{*}} g_{\psi D D^{*}} e \epsilon^{\beta \eta \rho \sigma} \epsilon_{1}^{\alpha} \epsilon^{*}_{2\eta} \epsilon^{* \mu}_{3}\right) p_{2\beta} \left( p_2 - q_1 \right)_{\rho} \left\lbrace \left( -g_{\sigma}^{\nu} + \frac{\left(p_2 - q_1\right)_{\sigma}\left(p_2-q_1\right)^{\nu}}{m_{D^{*}}^{2}} \right) \left[ \left( -g_{\alpha\nu} + \frac{\left(p_1-q_1\right)_{\alpha}\left(p_1-q_1\right)_{\nu}}{m_{D^{*}}^{2}} \right) \right. \right. \nonumber \\
			&\left. \left(p_1-q_1\right)_{\mu} + \left(-g_{\alpha\mu}+\frac{\left(p_1-q_1\right)_{\alpha}\left(p_1-q_1\right)_{\mu}}{m_{D^{*}}^{2}}\right)\left(p_1-q_1\right)_{\nu} \right] - \left(-g_{\alpha}^{\nu} + \frac{\left(p_1-q_1\right)_{\alpha}\left(p_1-q_1\right)^{\nu}}{m_{D^{*}}^{2}} \right) \left[ \left(-g_{\sigma\nu}+\frac{\left(p_2-q_1\right)_{\sigma}\left(p_2-q_1\right)_{\nu}}{m_{D^{*}}^{2}}\right) \right. \nonumber \\
			&\left. \left. \left(p_2-q_1\right)_{\mu} + \left(-g_{\sigma\mu}+\frac{\left(p_2-q_1\right)_{\sigma}\left(p_2-q_1\right)_{\mu}}{m_{D^{*}}^{2}}\right) \left(p_2-q_1\right)_{\nu} \right] \right\rbrace \left(\frac{1}{q_{1}^{2}-m_{D}^{2}}\right)  \left(\frac{1}{\left(p_1 - q_1 \right)^{2}-m_{D^{*}}^{2}}\right) \left(\frac{1}{\left(p_2 - q_1 \right)^{2}-m_{D^{*}}^{2}}\right) \nonumber \\
			&\times  F(m_{D},q_{1}^{2}) F(m_{D^{*}},\left(p_{1}-q_{1}\right)^{2}) F(m_{D^{*}},\left(p_{2}-q_{1}\right)^{2}), \nonumber \\
			\mathcal{M}_{c} &= \int \frac{d^{4}q_1}{\left( 2 \pi \right)^{4}} \left( g_{\chi D D^{*}} g_{\psi D^{*}D^{*}} e g_{\gamma D D^{*}} \epsilon^{\mu \nu \alpha \beta} \right) \epsilon_{1}^{\rho} \epsilon^{*}_{3 \nu} p_{3\mu}\left( p_2 - q_1 \right)_{\alpha} \left\lbrace \epsilon_{2}^{*\sigma}\left(p_2-2q_1\right)_{\sigma} \left(-g_{\rho}^{\eta}+\frac{q_{1 \rho}q_{1}^{\eta}}{m_{D^{*}}^{2}}\right)\left(-g_{\eta \beta} + \frac{\left(p_{2}-q_{1}\right)_{\eta}\left(p_{2}-q_{1}\right)_{\beta}}{m_{D^{*}}^{2}}\right)  \right. \nonumber \\
			&+ \left. \epsilon_{2 \eta}^{*}q_{1 \sigma} \left(-g_{\rho}^{\eta}+\frac{q_{1 \rho}q_{1}^{\eta}}{m_{D^{*}}^{2}}\right)\left(-g_{\beta}^{\sigma} + \frac{\left(p_{2}-q_{1}\right)_{\beta}\left(p_{2}-q_{1}\right)^{\sigma}}{m_{D^{*}}^{2}}\right) - \epsilon_{2}^{* \eta} \left(p_{2}-q_{1}\right)_{\sigma} \left(-g_{\rho}^{\sigma}+\frac{q_{1 \rho}q_{1}^{\sigma}}{m_{D^{*}}^{2}}\right) \left(-g_{\eta \beta} + \frac{\left(p_{2}-q_{1}\right)_{\eta}\left(p_{2}-q_{1}\right)_{\beta}}{m_{D^{*}}^{2}}\right) \right\rbrace \nonumber \\
			&\times \left(\frac{1}{q_{1}^{2}-m_{D^{*}}^{2}}\right)  \left(\frac{1}{\left(p_1 - q_1 \right)^{2}-m_{D}^{2}}\right) \left(\frac{1}{\left(p_2 - q_1 \right)^{2}-m_{D^{*}}^{2}}\right) F(m_{D^{*}},q_{1}^{2}) F(m_{D},\left(p_{1}-q_{1}\right)^{2}) F(m_{D^{*}},\left(p_{2}-q_{1}\right)^{2}),
		\end{align} 
	\end{widetext}
	
	where $\epsilon_{1}$, $\epsilon_{2}$, and $\epsilon_{3}$ are the polarization vectors of $\chi_{c1}$, $J/\psi$, and $\gamma$, respectively.  
	The momenta of $\bar{D}$ and $\bar{D}^*$ mesons in the triangle loop are denoted as $q_1$.
	The average masses of the $D$ and $D^{*}$ mesons are denoted as $m_{D}$ and $m_{D^{*}}$.

	The total amplitude for the radiative decay mode $\chi_{c1} \to J/\psi \gamma$ is given by
	\begin{equation}
		\mathcal{M} = \mathcal{M}_{a} + \mathcal{M}_{b} + \mathcal{M}_{c}.
	\end{equation}

	The branching fraction for the $i$th decay mode is then calculated as
	\begin{equation}
		R_{i} = \frac{\Gamma_{i}}{\Gamma},
	\end{equation}
	with
	\begin{align}
		\Gamma_{i} &= \frac{\left| \vec{q} \right|}{32 \pi^{2} m_{\chi}^{2}} \int \left\langle \left| \mathcal{M} \right|^{2} \right\rangle d\Omega, \\
		\left\langle \left| \mathcal{M} \right|^{2} \right\rangle &= \frac{1}{3} \sum_{\lambda_{1},\lambda_{2},\lambda_{3}} \left| \mathcal{M} \right|^{2}.
	\end{align}
	
	\section{Results}
	\label{result}
	
	First, we fix the free parameters $\alpha_{D}$ and $\alpha_{D^{*}}$ such that the observed branching fraction of the decay mode $\chi_{c1}(1P) \to J/\psi \gamma$ is reproduced. 
	The values of the observed branching fraction $R^{(e)}_{\chi_{c1} \to J/ \psi \gamma}$ and our result $R^{(t)}_{\chi_{c1} \to J/ \psi \gamma}$ are
	\begin{align}
		R^{(e)}_{\chi_{c1} \to J/ \psi \gamma} &= 0.343 \pm 0.013 \nonumber, \\
		R^{(t)}_{\chi_{c1} \to J/ \psi \gamma} &= 0.357 \pm 0.017,
	\end{align}
	where the parameters $\alpha_{D}$ and $\alpha_{D^{*}}$ are set to
	\begin{align}
		\alpha_{D} &= 1.65, \nonumber\\
		\alpha_{D^*} &= 1.01.
	\end{align}
	By using these parameters, we predict the fractions for the radiative decay modes $\chi_{c1}(3872) \to J/\psi \gamma$ and $\chi_{c1}(3872) \to \psi(2S) \gamma$. 
	The values of the observed branching fractions $R^{(e)}$ (if available) and our predictions $R^{(t)}$ are summarized in Table. \ref{prediction}.
	
\begin{center}
	\begin{table}[h]
	\begin{tabular}[c]{|c|c|c|}
		\hline
		Decay mode & $R^{(e)}$ & $R^{(t)}$ \\
		\hline
		$\chi_{c1}(3872) \to J/\psi \gamma$ & $\left(10 \pm 4 \right) \times 10^{-3}$ & $\left(3.2 \pm 0.6 \right) \times 10^{-1}$ \\
		$\chi_{c1}(3872) \to \psi(2S) \gamma$ & - & $\left(3.5 \pm 0.6 \right) \times 10^{-2}$ \\
		\hline
	\end{tabular}
		\caption{Radiative decay fractions of $\chi_{c1}(3872)$}
		\label{prediction}
\end{table}
\end{center}
	In comparison to the observed fraction, our prediction for the decay channel $\chi_{c1}(3872) \to J/\psi \gamma$ is about two order of magnitude above the data. 
	In addition, we also consider the relative ratio for the radiative decay fractions $\mathcal{R}_{\Psi \gamma} = \frac{R_{\chi_{c1}(3872) \to \psi(2S) \gamma}}{R_{\chi_{c1}(3872) \to J/\psi \gamma}}$.
	The observed data at LHCb \cite{LHCb:2024tpv} and result from our study are
	\begin{align}
		\mathcal{R}^{(e)}_{\Psi \gamma} &= 1.67 \pm 0.21 \pm 0.12 \pm 0.04, \nonumber \\
		\mathcal{R}^{(t)}_{\Psi \gamma} &= 0.109 \pm 0.028,
	\end{align}
	where our prediction is about one order of magnitude below the observed value of $\mathcal{R}_{\Psi\gamma}$.
	From these analyses, the internal structure of $\chi_{c1}(3872)$ may differ from that of the radial excitation $\chi_{c1}(2P)$ state.

	\section{Summary}
	\label{sum}
	In this study, we have investigated the radiative decay of $\chi_{c1}$ states using the effective Lagrangian approach, incorporating the triangle loops of $D$ and $D^{*}$ mesons.  
	The coupling constants are taken from several sources, while the cutoff parameters for the form factors of the $D$ and $D^{*}$ intermediate states are parameterized in terms of the free parameter $\alpha$.
	To reproduce the observed fraction of $\chi_{c1}(1P) \to J/\psi \gamma$, the free parameters $\alpha_{D}$ and $\alpha_{D^{*}}$ are set to 1.97 and 1.32, respectively.
	Then, the same set of parameters is employed to investigate the fractions of the decay channels $\chi_{c1}(3872) \to J/\psi \gamma$, $\psi(2S)\gamma$.
	The branching fractions are predicted to be $R_{\chi_{c1}(3872) \to J/\psi \gamma} \sim 10^{-1}$ and $R_{\chi_{c1}(3872) \to \psi(2S) \gamma} \sim 10^{-2}$, with the relative ratio given by $\mathcal{R}_{\Psi \gamma} \approx 0.109$.
	
	From our analyses, the fraction $R_{\chi_{c1}(3872) \to J/\psi \gamma}$ is predicted to be about two orders of magnitude above the experimental value.
	Additionally, the predicted relative fraction $R_{\Psi\gamma}$ is about one order of magnitude lower than the observed value reported by LHCb.
	These discrepancies suggest that the internal structure of the exotic $\chi_{c1}(3872)$ state may differ from that of the radial excitation $\chi_{c1}(2P)$ state.
	
	\begin{acknowledgments}
	This research has received funding support from the NSRF via the Program Management Unit for Human Resources \& Institutional Development, Research and Innovation [grant number B50G670107].
	\end{acknowledgments}


\begin{thebibliography}{}
		
	%experiment
		
	%\cite{Belle:2003nnu}
	\bibitem{Belle:2003nnu}
	S.~K.~Choi \textit{et al.} [Belle],
	%``Observation of a narrow charmonium-like state in exclusive $B^\pm \to K^\pm \pi^+ \pi^- J/\psi$ decays,''
	Phys. Rev. Lett. \textbf{91}, 262001 (2003)
	doi:10.1103/PhysRevLett.91.262001
	[arXiv:hep-ex/0309032 [hep-ex]].
	%2654 citations counted in INSPIRE as of 23 Jan 2025
	
	%\cite{Belle:2003eeg}
	\bibitem{Belle:2003eeg}
	K.~Abe \textit{et al.} [Belle],
	%``Observation of B+ ---\ensuremath{>} psi(3770) K+,''
	Phys. Rev. Lett. \textbf{93}, 051803 (2004)
	doi:10.1103/PhysRevLett.93.051803
	[arXiv:hep-ex/0307061 [hep-ex]].
	%73 citations counted in INSPIRE as of 23 Jan 2025
	
	%\cite{Belle:2006olv}
	\bibitem{Belle:2006olv}
	G.~Gokhroo \textit{et al.} [Belle],
	%``Observation of a Near-threshold D0 anti-D0 pi0 Enhancement in B ---\ensuremath{>} D0 anti-D0 pi0 K Decay,''
	Phys. Rev. Lett. \textbf{97}, 162002 (2006)
	doi:10.1103/PhysRevLett.97.162002
	[arXiv:hep-ex/0606055 [hep-ex]].
	%284 citations counted in INSPIRE as of 23 Jan 2025
	
	%\cite{Belle:2008fma}
	\bibitem{Belle:2008fma}
	T.~Aushev \textit{et al.} [Belle],
	%``Study of the B ---\ensuremath{>} X(3872)(D*0 anti-D0) K decay,''
	Phys. Rev. D \textbf{81}, 031103 (2010)
	doi:10.1103/PhysRevD.81.031103
	[arXiv:0810.0358 [hep-ex]].
	%217 citations counted in INSPIRE as of 23 Jan 2025
	
	%\cite{Belle:2011vlx}
	\bibitem{Belle:2011vlx}
	S.~K.~Choi \textit{et al.} [Belle],
	%``Bounds on the width, mass difference and other properties of $X(3872) \to \pi^+ \pi^- J/\psi$ decays,''
	Phys. Rev. D \textbf{84}, 052004 (2011)
	doi:10.1103/PhysRevD.84.052004
	[arXiv:1107.0163 [hep-ex]].
	%264 citations counted in INSPIRE as of 23 Jan 2025
	
	%\cite{Belle:2011wdj}
	\bibitem{Belle:2011wdj}
	V.~Bhardwaj \textit{et al.} [Belle],
	%``Observation of $X(3872)\to J/\psi \gamma$ and search for $X(3872)\to\psi'\gamma$ in B decays,''
	Phys. Rev. Lett. \textbf{107}, 091803 (2011)
	doi:10.1103/PhysRevLett.107.091803
	[arXiv:1105.0177 [hep-ex]].
	%216 citations counted in INSPIRE as of 23 Jan 2025
	
	%\cite{Belle:2015qeg}
	\bibitem{Belle:2015qeg}
	A.~Bala \textit{et al.} [Belle],
	%``Observation of X(3872) in B\textrightarrow{}X(3872)K\ensuremath{\pi} decays,''
	Phys. Rev. D \textbf{91}, no.5, 051101 (2015)
	doi:10.1103/PhysRevD.91.051101
	[arXiv:1501.06867 [hep-ex]].
	%36 citations counted in INSPIRE as of 23 Jan 2025
	
	%\cite{Belle:2022puc}
	\bibitem{Belle:2022puc}
	J.~H.~Yin \textit{et al.} [Belle],
	%``Search for X(3872)\textrightarrow{}\ensuremath{\pi}+\ensuremath{\pi}-\ensuremath{\pi}0 at Belle,''
	Phys. Rev. D \textbf{107}, no.5, 052004 (2023)
	doi:10.1103/PhysRevD.107.052004
	[arXiv:2206.08592 [hep-ex]].
	%6 citations counted in INSPIRE as of 23 Jan 2025
	
	%\cite{Belle:2023zxm}
	\bibitem{Belle:2023zxm}
	H.~Hirata \textit{et al.} [Belle],
	%``Study of the lineshape of X(3872) using B decays to D0D\textasciimacron{}*0K,''
	Phys. Rev. D \textbf{107}, no.11, 112011 (2023)
	doi:10.1103/PhysRevD.107.112011
	[arXiv:2302.02127 [hep-ex]].
	%14 citations counted in INSPIRE as of 23 Jan 2025
	
	%\cite{BaBar:2004iez}
	\bibitem{BaBar:2004iez}
	B.~Aubert \textit{et al.} [BaBar],
	%``Observation of the decay $B \to J/\psi \eta K$ and search for $X(3872) \to J/\psi \eta$,''
	Phys. Rev. Lett. \textbf{93}, 041801 (2004)
	doi:10.1103/PhysRevLett.93.041801
	[arXiv:hep-ex/0402025 [hep-ex]].
	%114 citations counted in INSPIRE as of 23 Jan 2025
	
	%\cite{BaBar:2004oro}
	\bibitem{BaBar:2004oro}
	B.~Aubert \textit{et al.} [BaBar],
	%``Study of the $B \to J/\psi K^- \pi^+ \pi^-$ decay and measurement of the $B \to X(3872) K^-$ branching fraction,''
	Phys. Rev. D \textbf{71}, 071103 (2005)
	doi:10.1103/PhysRevD.71.071103
	[arXiv:hep-ex/0406022 [hep-ex]].
	%766 citations counted in INSPIRE as of 23 Jan 2025
	
	%\cite{BaBar:2006fjg}
	\bibitem{BaBar:2006fjg}
	B.~Aubert \textit{et al.} [BaBar],
	%``Search for $B^{+} \to X(3872) K^{+}$, $X_{3872} \to J/\psi \gamma$,''
	Phys. Rev. D \textbf{74}, 071101 (2006)
	doi:10.1103/PhysRevD.74.071101
	[arXiv:hep-ex/0607050 [hep-ex]].
	%221 citations counted in INSPIRE as of 23 Jan 2025
	
	%\cite{BaBar:2007ixp}
	\bibitem{BaBar:2007ixp}
	B.~Aubert \textit{et al.} [BaBar],
	%``Search for Prompt Production of chi(c) and X(3872) in e+e- Annihilations,''
	Phys. Rev. D \textbf{76}, 071102 (2007)
	doi:10.1103/PhysRevD.76.071102
	[arXiv:0707.1633 [hep-ex]].
	%9 citations counted in INSPIRE as of 23 Jan 2025
	
	%\cite{BaBar:2007cmo}
	\bibitem{BaBar:2007cmo}
	B.~Aubert \textit{et al.} [BaBar],
	%``Study of Resonances in Exclusive B Decays to anti-D(*) D(*) K,''
	Phys. Rev. D \textbf{77}, 011102 (2008)
	doi:10.1103/PhysRevD.77.011102
	[arXiv:0708.1565 [hep-ex]].
	%272 citations counted in INSPIRE as of 23 Jan 2025
	
	%\cite{BaBar:2008qzi}
	\bibitem{BaBar:2008qzi}
	B.~Aubert \textit{et al.} [BaBar],
	%``A Study of $B \to X(3872) K$, with $X_{3872} \to J/\Psi \pi^{+} \pi^{-}$,''
	Phys. Rev. D \textbf{77}, 111101 (2008)
	doi:10.1103/PhysRevD.77.111101
	[arXiv:0803.2838 [hep-ex]].
	%186 citations counted in INSPIRE as of 23 Jan 2025
	
	%\cite{BaBar:2008flx}
	\bibitem{BaBar:2008flx}
	B.~Aubert \textit{et al.} [BaBar],
	%``Evidence for $X(3872) \to \psi_{2S} \gamma$ in $B^\pm \to X_{3872} K^\pm$ decays, and a study of $B \to c \bar{c} \gamma K$,''
	Phys. Rev. Lett. \textbf{102}, 132001 (2009)
	doi:10.1103/PhysRevLett.102.132001
	[arXiv:0809.0042 [hep-ex]].
	%325 citations counted in INSPIRE as of 23 Jan 2025
	
	%\cite{BaBar:2010wfc}
	\bibitem{BaBar:2010wfc}
	P.~del Amo Sanchez \textit{et al.} [BaBar],
	%``Evidence for the decay X(3872) ---\ensuremath{>} J/ psi omega,''
	Phys. Rev. D \textbf{82}, 011101 (2010)
	doi:10.1103/PhysRevD.82.011101
	[arXiv:1005.5190 [hep-ex]].
	%320 citations counted in INSPIRE as of 23 Jan 2025
	
	%\cite{CDF:2003cab}
	\bibitem{CDF:2003cab}
	D.~Acosta \textit{et al.} [CDF],
	%``Observation of the narrow state $X(3872) \to J/\psi \pi^+ \pi^-$ in $\bar{p}p$ collisions at $\sqrt{s} = 1.96$ TeV,''
	Phys. Rev. Lett. \textbf{93}, 072001 (2004)
	doi:10.1103/PhysRevLett.93.072001
	[arXiv:hep-ex/0312021 [hep-ex]].
	%974 citations counted in INSPIRE as of 23 Jan 2025
	
	%\cite{CDF:2005cfq}
	\bibitem{CDF:2005cfq}
	A.~Abulencia \textit{et al.} [CDF],
	%``Measurement of the dipion mass spectrum in $X(3872) \to J/\psi \pi^+ \pi^-$ decays.,''
	Phys. Rev. Lett. \textbf{96}, 102002 (2006)
	doi:10.1103/PhysRevLett.96.102002
	[arXiv:hep-ex/0512074 [hep-ex]].
	%215 citations counted in INSPIRE as of 23 Jan 2025
	
	%\cite{CDF:2006ocq}
	\bibitem{CDF:2006ocq}
	A.~Abulencia \textit{et al.} [CDF],
	%``Analysis of the quantum numbers $J^{PC}$ of the $X(3872)$,''
	Phys. Rev. Lett. \textbf{98}, 132002 (2007)
	doi:10.1103/PhysRevLett.98.132002
	[arXiv:hep-ex/0612053 [hep-ex]].
	%371 citations counted in INSPIRE as of 23 Jan 2025
	
	%\cite{CDF:2009nxk}
	\bibitem{CDF:2009nxk}
	T.~Aaltonen \textit{et al.} [CDF],
	%``Precision Measurement of the $X(3872)$ Mass in $J/\psi \pi^+ \pi^-$ Decays,''
	Phys. Rev. Lett. \textbf{103}, 152001 (2009)
	doi:10.1103/PhysRevLett.103.152001
	[arXiv:0906.5218 [hep-ex]].
	%177 citations counted in INSPIRE as of 23 Jan 2025
	
	%\cite{D0:2004zmu}
	\bibitem{D0:2004zmu}
	V.~M.~Abazov \textit{et al.} [D0],
	%``Observation and properties of the $X(3872)$ decaying to $J/\psi \pi^+ \pi^-$ in $p\bar{p}$ collisions at $\sqrt{s} = 1.96$ TeV,''
	Phys. Rev. Lett. \textbf{93}, 162002 (2004)
	doi:10.1103/PhysRevLett.93.162002
	[arXiv:hep-ex/0405004 [hep-ex]].
	%854 citations counted in INSPIRE as of 23 Jan 2025
	
	%\cite{ATLAS:2016kwu}
	\bibitem{ATLAS:2016kwu}
	M.~Aaboud \textit{et al.} [ATLAS],
	%``Measurements of $\psi(2S)$ and $X(3872) \to J/\psi\pi^+\pi^-$ production in $pp$ collisions at $\sqrt{s} = 8$ TeV with the ATLAS detector,''
	JHEP \textbf{01}, 117 (2017)
	doi:10.1007/JHEP01(2017)117
	[arXiv:1610.09303 [hep-ex]].
	%111 citations counted in INSPIRE as of 23 Jan 2025
	
	%\cite{CMS:2013fpt}
	\bibitem{CMS:2013fpt}
	S.~Chatrchyan \textit{et al.} [CMS],
	%``Measurement of the $X$(3872) Production Cross Section Via Decays to $J/\psi \pi^+ \pi^-$ in $pp$ collisions at $\sqrt{s}$ = 7 TeV,''
	JHEP \textbf{04}, 154 (2013)
	doi:10.1007/JHEP04(2013)154
	[arXiv:1302.3968 [hep-ex]].
	%271 citations counted in INSPIRE as of 23 Jan 2025
	
	%\cite{CMS:2020eiw}
	\bibitem{CMS:2020eiw}
	A.~M.~Sirunyan \textit{et al.} [CMS],
	%``Observation of the B$^0_\mathrm{s}\to $X(3872)$\phi$ decay,''
	Phys. Rev. Lett. \textbf{125}, no.15, 152001 (2020)
	doi:10.1103/PhysRevLett.125.152001
	[arXiv:2005.04764 [hep-ex]].
	%44 citations counted in INSPIRE as of 23 Jan 2025
	
	%\cite{LHCb:2011zzp}
	\bibitem{LHCb:2011zzp}
	R.~Aaij \textit{et al.} [LHCb],
	%``Observation of $X(3872) $ production in $pp$ collisions at $\sqrt{s}=7$ TeV,''
	Eur. Phys. J. C \textbf{72}, 1972 (2012)
	doi:10.1140/epjc/s10052-012-1972-7
	[arXiv:1112.5310 [hep-ex]].
	%280 citations counted in INSPIRE as of 23 Jan 2025
	
	%\cite{LHCb:2013kgk}
	\bibitem{LHCb:2013kgk}
	R.~Aaij \textit{et al.} [LHCb],
	%``Determination of the X(3872) meson quantum numbers,''
	Phys. Rev. Lett. \textbf{110}, 222001 (2013)
	doi:10.1103/PhysRevLett.110.222001
	[arXiv:1302.6269 [hep-ex]].
	%545 citations counted in INSPIRE as of 23 Jan 2025
	
	%\cite{LHCb:2014jvf}
	\bibitem{LHCb:2014jvf}
	R.~Aaij \textit{et al.} [LHCb],
	%``Evidence for the decay $X(3872)\rightarrow\psi(2S)\gamma$,''
	Nucl. Phys. B \textbf{886}, 665-680 (2014)
	doi:10.1016/j.nuclphysb.2014.06.011
	[arXiv:1404.0275 [hep-ex]].
	%187 citations counted in INSPIRE as of 23 Jan 2025
	
	%\cite{LHCb:2015jfc}
	\bibitem{LHCb:2015jfc}
	R.~Aaij \textit{et al.} [LHCb],
	%``Quantum numbers of the $X(3872)$ state and orbital angular momentum in its $\rho^0 J/\psi$ decay,''
	Phys. Rev. D \textbf{92}, no.1, 011102 (2015)
	doi:10.1103/PhysRevD.92.011102
	[arXiv:1504.06339 [hep-ex]].
	%150 citations counted in INSPIRE as of 23 Jan 2025
	
	%\cite{LHCb:2016zqv}
	\bibitem{LHCb:2016zqv}
	R.~Aaij \textit{et al.} [LHCb],
	%``Observation of $\eta_{c}(2S) \to p \bar p$ and search for $X(3872) \to p \bar p$ decays,''
	Phys. Lett. B \textbf{769}, 305-313 (2017)
	doi:10.1016/j.physletb.2017.03.046
	[arXiv:1607.06446 [hep-ex]].
	%39 citations counted in INSPIRE as of 23 Jan 2025
	
	%\cite{LHCb:2019imv}
	\bibitem{LHCb:2019imv}
	R.~Aaij \textit{et al.} [LHCb],
	%``Observation of the $\Lambda_b^0\rightarrow \chi_{c1}(3872)pK^-$ decay,''
	JHEP \textbf{09}, 028 (2019)
	doi:10.1007/JHEP09(2019)028
	[arXiv:1907.00954 [hep-ex]].
	%26 citations counted in INSPIRE as of 23 Jan 2025
	
	%\cite{LHCb:2020xds}
	\bibitem{LHCb:2020xds}
	R.~Aaij \textit{et al.} [LHCb],
	%``Study of the lineshape of the $\chi_{c1}(3872)$ state,''
	Phys. Rev. D \textbf{102}, no.9, 092005 (2020)
	doi:10.1103/PhysRevD.102.092005
	[arXiv:2005.13419 [hep-ex]].
	%154 citations counted in INSPIRE as of 23 Jan 2025
	
	%\cite{LHCb:2020fvo}
	\bibitem{LHCb:2020fvo}
	R.~Aaij \textit{et al.} [LHCb],
	%``Study of the $\psi_2(3823)$ and $\chi_{c1}(3872)$ states in $B^+ \rightarrow \left( J\psi\pi^+\pi^-\right)K^+$ decays,''
	JHEP \textbf{08}, 123 (2020)
	doi:10.1007/JHEP08(2020)123
	[arXiv:2005.13422 [hep-ex]].
	%105 citations counted in INSPIRE as of 23 Jan 2025
	
	%\cite{LHCb:2020sey}
	\bibitem{LHCb:2020sey}
	R.~Aaij \textit{et al.} [LHCb],
	%``Observation of Multiplicity Dependent Prompt $\chi_{c1}(3872)$ and $\psi(2S)$ Production in $pp$ Collisions,''
	Phys. Rev. Lett. \textbf{126}, no.9, 092001 (2021)
	doi:10.1103/PhysRevLett.126.092001
	[arXiv:2009.06619 [hep-ex]].
	%69 citations counted in INSPIRE as of 23 Jan 2025
	
	%\cite{LHCb:2021ten}
	\bibitem{LHCb:2021ten}
	R.~Aaij \textit{et al.} [LHCb],
	%``Measurement of \ensuremath{\chi}$_{c1}$(3872) production in proton-proton collisions at $ \sqrt{s} $ = 8 and 13 TeV,''
	JHEP \textbf{01}, 131 (2022)
	doi:10.1007/JHEP01(2022)131
	[arXiv:2109.07360 [hep-ex]].
	%32 citations counted in INSPIRE as of 23 Jan 2025
	
	%\cite{LHCb:2022jez}
	\bibitem{LHCb:2022jez}
	R.~Aaij \textit{et al.} [LHCb],
	%``Observation of sizeable \ensuremath{\omega} contribution to \ensuremath{\chi}c1(3872)\textrightarrow{}\ensuremath{\pi}+\ensuremath{\pi}-J/\ensuremath{\psi} decays,''
	Phys. Rev. D \textbf{108}, no.1, L011103 (2023)
	doi:10.1103/PhysRevD.108.L011103
	[arXiv:2204.12597 [hep-ex]].
	%35 citations counted in INSPIRE as of 23 Jan 2025
	
	%\cite{LHCb:2022oqs}
	\bibitem{LHCb:2022oqs}
	R.~Aaij \textit{et al.} [LHCb],
	%``Study of charmonium and charmonium-like contributions in $B^+ \rightarrow J/\psi \eta K^+$ decays,''
	JHEP \textbf{04}, 046 (2022)
	doi:10.1007/JHEP04(2022)046
	[arXiv:2202.04045 [hep-ex]].
	%11 citations counted in INSPIRE as of 23 Jan 2025
	
	%\cite{LHCb:2024bpb}
	\bibitem{LHCb:2024bpb}
	R.~Aaij \textit{et al.} [LHCb],
	%``Modification of \ensuremath{\chi}c1(3872) and \ensuremath{\psi}(2S) Production in pPb Collisions at sNN=8.16\,\,TeV,''
	Phys. Rev. Lett. \textbf{132}, no.24, 242301 (2024)
	doi:10.1103/PhysRevLett.132.242301
	[arXiv:2402.14975 [hep-ex]].
	%11 citations counted in INSPIRE as of 23 Jan 2025
	
	%charmonium
	
	%\cite{Barnes:2003vb}
	\bibitem{Barnes:2003vb}
	T.~Barnes and S.~Godfrey,
	%``Charmonium options for the X(3872),''
	Phys. Rev. D \textbf{69}, 054008 (2004)
	doi:10.1103/PhysRevD.69.054008
	[arXiv:hep-ph/0311162 [hep-ph]].
	%347 citations counted in INSPIRE as of 06 Feb 2025
	
	%\cite{Kalashnikova:2010hv}
	\bibitem{Kalashnikova:2010hv}
	Y.~S.~Kalashnikova and A.~V.~Nefediev,
	%``X(3872) as a 1D2 charmonium state,''
	Phys. Rev. D \textbf{82}, 097502 (2010)
	doi:10.1103/PhysRevD.82.097502
	[arXiv:1008.2895 [hep-ph]].
	%42 citations counted in INSPIRE as of 06 Feb 2025
	
	%\cite{Yu:2023nxk}
	\bibitem{Yu:2023nxk}
	S.~Y.~Yu and X.~W.~Kang,
	%``Nature of X(3872) from its radiative decay,''
	Phys. Lett. B \textbf{848}, 138404 (2024)
	doi:10.1016/j.physletb.2023.138404
	[arXiv:2308.10219 [hep-ph]].
	%9 citations counted in INSPIRE as of 06 Feb 2025
	
	%Molecule picture

	%\cite{Tornqvist:2004qy}
	\bibitem{Tornqvist:2004qy}
	N.~A.~Tornqvist,
	%``Isospin breaking of the narrow charmonium state of Belle at 3872-MeV as a deuson,''
	Phys. Lett. B \textbf{590}, 209-215 (2004)
	doi:10.1016/j.physletb.2004.03.077
	[arXiv:hep-ph/0402237 [hep-ph]].
	%625 citations counted in INSPIRE as of 06 Feb 2025
	
	%\cite{Liu:2008fh}
	\bibitem{Liu:2008fh}
	Y.~R.~Liu, X.~Liu, W.~Z.~Deng and S.~L.~Zhu,
	%``Is $X(3872) $ Really a Molecular State?,''
	Eur. Phys. J. C \textbf{56}, 63-73 (2008)
	doi:10.1140/epjc/s10052-008-0640-4
	[arXiv:0801.3540 [hep-ph]].
	%185 citations counted in INSPIRE as of 06 Feb 2025
	
	%\cite{Liu:2009qhy}
	\bibitem{Liu:2009qhy}
	X.~Liu, Z.~G.~Luo, Y.~R.~Liu and S.~L.~Zhu,
	%``X(3872) and Other Possible Heavy Molecular States,''
	Eur. Phys. J. C \textbf{61}, 411-428 (2009)
	doi:10.1140/epjc/s10052-009-1020-4
	[arXiv:0808.0073 [hep-ph]].
	%241 citations counted in INSPIRE as of 06 Feb 2025
	
	%\cite{Li:2012cs}
	\bibitem{Li:2012cs}
	N.~Li and S.~L.~Zhu,
	%``Isospin breaking, Coupled-channel effects and Diagnosis of X(3872),''
	Phys. Rev. D \textbf{86}, 074022 (2012)
	doi:10.1103/PhysRevD.86.074022
	[arXiv:1207.3954 [hep-ph]].
	%82 citations counted in INSPIRE as of 06 Feb 2025
	
	%\cite{Guo:2013zbw}
	\bibitem{Guo:2013zbw}
	F.~K.~Guo, C.~Hanhart, U.~G.~Mei\ss{}ner, Q.~Wang and Q.~Zhao,
	%``Production of the X(3872) in charmonia radiative decays,''
	Phys. Lett. B \textbf{725}, 127-133 (2013)
	doi:10.1016/j.physletb.2013.06.053
	[arXiv:1306.3096 [hep-ph]].
	%134 citations counted in INSPIRE as of 06 Feb 2025
	
	%\cite{Guo:2014taa}
	\bibitem{Guo:2014taa}
	F.~K.~Guo, C.~Hanhart, Y.~S.~Kalashnikova, U.~G.~Mei\ss{}ner and A.~V.~Nefediev,
	%``What can radiative decays of the X(3872) teach us about its nature?,''
	Phys. Lett. B \textbf{742}, 394-398 (2015)
	doi:10.1016/j.physletb.2015.02.013
	[arXiv:1410.6712 [hep-ph]].
	%68 citations counted in INSPIRE as of 06 Feb 2025
	
	%\cite{Chen:2024xlw}
	\bibitem{Chen:2024xlw}
	P.~Chen, Z.~W.~Liu, Z.~L.~Zhang, S.~Q.~Luo, F.~L.~Wang, J.~Z.~Wang and X.~Liu,
	%``Role of electromagnetic interactions in the X(3872) and its analogs,''
	Phys. Rev. D \textbf{109}, no.9, 094002 (2024)
	doi:10.1103/PhysRevD.109.094002
	[arXiv:2401.05989 [hep-ph]].
	%4 citations counted in INSPIRE as of 06 Feb 2025
	
	%Couple channel
	
	%\cite{Braaten:2003he}
	\bibitem{Braaten:2003he}
	E.~Braaten and M.~Kusunoki,
	%``Low-energy universality and the new charmonium resonance at 3870-MeV,''
	Phys. Rev. D \textbf{69}, 074005 (2004)
	doi:10.1103/PhysRevD.69.074005
	[arXiv:hep-ph/0311147 [hep-ph]].
	%304 citations counted in INSPIRE as of 07 Feb 2025
	
	%\cite{Kalashnikova:2005ui}
	\bibitem{Kalashnikova:2005ui}
	Y.~S.~Kalashnikova,
	%``Coupled-channel model for charmonium levels and an option for X(3872),''
	Phys. Rev. D \textbf{72}, 034010 (2005)
	doi:10.1103/PhysRevD.72.034010
	[arXiv:hep-ph/0506270 [hep-ph]].
	%208 citations counted in INSPIRE as of 07 Feb 2025
	
	%\cite{Barnes:2007xu}
	\bibitem{Barnes:2007xu}
	T.~Barnes and E.~S.~Swanson,
	%``Hadron loops: General theorems and application to charmonium,''
	Phys. Rev. C \textbf{77}, 055206 (2008)
	doi:10.1103/PhysRevC.77.055206
	[arXiv:0711.2080 [hep-ph]].
	%149 citations counted in INSPIRE as of 07 Feb 2025
		
	%\cite{Li:2009ad}
	\bibitem{Li:2009ad}
	B.~Q.~Li, C.~Meng and K.~T.~Chao,
	%``Coupled-Channel and Screening Effects in Charmonium Spectrum,''
	Phys. Rev. D \textbf{80}, 014012 (2009)
	doi:10.1103/PhysRevD.80.014012
	[arXiv:0904.4068 [hep-ph]].
	%134 citations counted in INSPIRE as of 07 Feb 2025
	
	%\cite{Ortega:2009hj}
	\bibitem{Ortega:2009hj}
	P.~G.~Ortega, J.~Segovia, D.~R.~Entem and F.~Fernandez,
	%``Coupled channel approach to the structure of the X(3872),''
	Phys. Rev. D \textbf{81}, 054023 (2010)
	doi:10.1103/PhysRevD.81.054023
	[arXiv:0907.3997 [hep-ph]].
	%147 citations counted in INSPIRE as of 07 Feb 2025
	
	%\cite{Baru:2010ww}
	\bibitem{Baru:2010ww}
	V.~Baru, C.~Hanhart, Y.~S.~Kalashnikova, A.~E.~Kudryavtsev and A.~V.~Nefediev,
	%``Interplay of quark and meson degrees of freedom in a near-threshold resonance,''
	Eur. Phys. J. A \textbf{44}, 93-103 (2010)
	doi:10.1140/epja/i2010-10929-7
	[arXiv:1001.0369 [hep-ph]].
	%86 citations counted in INSPIRE as of 06 Feb 2025
	
	%\cite{Cardoso:2014xda}
	\bibitem{Cardoso:2014xda}
	M.~Cardoso, G.~Rupp and E.~van Beveren,
	%``Unquenched quark-model calculation of $X(3872)$ electromagnetic decays,''
	Eur. Phys. J. C \textbf{75}, no.1, 26 (2015)
	doi:10.1140/epjc/s10052-014-3254-z
	[arXiv:1411.1654 [hep-ph]].
	%27 citations counted in INSPIRE as of 06 Feb 2025
	
	%\cite{Yamaguchi:2019vea}
	\bibitem{Yamaguchi:2019vea}
	Y.~Yamaguchi, A.~Hosaka, S.~Takeuchi and M.~Takizawa,
	%``Heavy hadronic molecules with pion exchange and quark core couplings: a guide for practitioners,''
	J. Phys. G \textbf{47}, no.5, 053001 (2020)
	doi:10.1088/1361-6471/ab72b0
	[arXiv:1908.08790 [hep-ph]].
	%76 citations counted in INSPIRE as of 06 Feb 2025
	
	%\cite{Takeuchi:2021cnp}
	\bibitem{Takeuchi:2021cnp}
	S.~Takeuchi, Y.~Yamaguchi, A.~Hosaka and M.~Takizawa,
	%``X(3872) Revisited: The Roles of OPEP and the Quark Degrees of Freedom,''
	Few Body Syst. \textbf{62}, no.4, 77 (2021)
	doi:10.1007/s00601-021-01663-8
	[arXiv:2405.16794 [hep-ph]].
	%3 citations counted in INSPIRE as of 06 Feb 2025
	
	%compact tetraquark
	
	%\cite{Maiani:2004vq}
	\bibitem{Maiani:2004vq}
	L.~Maiani, F.~Piccinini, A.~D.~Polosa and V.~Riquer,
	%``Diquark-antidiquarks with hidden or open charm and the nature of X(3872),''
	Phys. Rev. D \textbf{71}, 014028 (2005)
	doi:10.1103/PhysRevD.71.014028
	[arXiv:hep-ph/0412098 [hep-ph]].
	%941 citations counted in INSPIRE as of 07 Feb 2025
	
	%\cite{Maiani:2007vr}
	\bibitem{Maiani:2007vr}
	L.~Maiani, A.~D.~Polosa and V.~Riquer,
	%``Indications of a Four-Quark Structure for the X(3872) and X(3876) Particles from Recent Belle and BABAR Data,''
	Phys. Rev. Lett. \textbf{99}, 182003 (2007)
	doi:10.1103/PhysRevLett.99.182003
	[arXiv:0707.3354 [hep-ph]].
	%103 citations counted in INSPIRE as of 07 Feb 2025
	
	%experiment
	
	%\cite{Accardi:2023chb}
	\bibitem{Accardi:2023chb}
	A.~Accardi, P.~Achenbach, D.~Adhikari, A.~Afanasev, C.~S.~Akondi, N.~Akopov, M.~Albaladejo, H.~Albataineh, M.~Albrecht and B.~Almeida-Zamora, \textit{et al.}
	%``Strong interaction physics at the luminosity frontier with 22 GeV electrons at Jefferson Lab,''
	Eur. Phys. J. A \textbf{60}, no.9, 173 (2024)
	doi:10.1140/epja/s10050-024-01282-x
	[arXiv:2306.09360 [nucl-ex]].
	%104 citations counted in INSPIRE as of 13 Feb 2025
	
	%Lagrangians

	%\cite{Meng:2007cx}
	\bibitem{Meng:2007cx}
	C.~Meng and K.~T.~Chao,
	%``Decays of the $X(3872) $ and chi(c1) (2P) charmonium,''
	Phys. Rev. D \textbf{75}, 114002 (2007)
	doi:10.1103/PhysRevD.75.114002
	[arXiv:hep-ph/0703205 [hep-ph]].
	%101 citations counted in INSPIRE as of 13 Feb 2025
	
	%\cite{Dong:2008gb}
	\bibitem{Dong:2008gb}
	Y.~b.~Dong, A.~Faessler, T.~Gutsche and V.~E.~Lyubovitskij,
	%``Estimate for the X(3872) ---\ensuremath{>} gamma J/psi decay width,''
	Phys. Rev. D \textbf{77}, 094013 (2008)
	doi:10.1103/PhysRevD.77.094013
	[arXiv:0802.3610 [hep-ph]].
	%160 citations counted in INSPIRE as of 13 Feb 2025
	
	%\cite{Liu:2006df}
	\bibitem{Liu:2006df}
	X.~Liu, B.~Zhang and S.~L.~Zhu,
	%``The Hidden Charm Decay of X(3872), Y(3940) and Final State Interaction Effects,''
	Phys. Lett. B \textbf{645}, 185-188 (2007)
	doi:10.1016/j.physletb.2006.12.031
	[arXiv:hep-ph/0610278 [hep-ph]].
	%73 citations counted in INSPIRE as of 13 Feb 2025
	
	%\cite{LHCb:2024tpv}
	\bibitem{LHCb:2024tpv}
	I.~Bezshyiko \textit{et al.} [LHCb],
	%``Probing the nature of the \ensuremath{\chi}$_{c1}$(3872) state using radiative decays,''
	JHEP \textbf{11}, 121 (2024)
	doi:10.1007/JHEP11(2024)121
	[arXiv:2406.17006 [hep-ex]].
	%21 citations counted in INSPIRE as of 04 Apr 2025
	
	\end{thebibliography}
\end{document}